# Measuring uncertainty during respiratory rate estimation using pressure-sensitive mats

S. Nizami, *MIEEE*, A. Bekele, M. Hozayen, K. Greenwood, J. Harrold and J. R. Green, *SMIEEE*

*Abstract*—We develop and evaluate a respiratory rate estimation algorithm that utilizes data from pressure-sensitive mat (PSM) technology for continuous patient monitoring in neonatal intensive care units (NICU). An analysis of the random effect of drift and systematic effect of creep in the PSM data is presented, showing that these are essentially dependent on the applied load and contact surface. Uncertainty measurements are pivotal when estimating physiologic parameters. The standard uncertainty in the PSM data is here represented by the percent drift. Next, we evaluate the applicability of PSM technology to estimate RR in neonatal patient simulator trials under five mixed effects including internally and externally induced motion, mattress type, grunting, laying position, and different breathing rates. We analyze the limits of agreement on the mixed effects model to derive the uncertainty in the estimated RR obtained through two estimation techniques. In comparison with the gold standard RR values, we achieved a mean bias of 0.56 breaths per minute (bpm) with an error bounded by a 95% confidence interval of [-2.26, 3.37] bpm. These results meet the clinical accuracy requirements of RR within ±5 bpm.

## I. INTRODUCTION

For several decades, continuous non-invasive respiratory monitoring has been used for early detection of life-threatening complications to improve patient outcomes [1]. Unexpected changes in respiratory rate (RR) may be indicators of serious illness [2], [3]. Yet, state-of-the-art respiratory monitoring technologies, including thoracic impedance electrode measurements commonly used in clinical practice, are known to be unreliable [4], [5]. Most RR errors are attributable to body movement artifacts [2], [4], [5]. More recently, contactless respiratory monitoring devices have made physiological data collection easier, especially from critically ill babies; however, methodological studies are needed to evaluate and improve the accuracy of these devices for critical care use [6]–[8]. For example, when a baby is breathing normally at 50-80 breaths/minute (bpm), then an accuracy within ±(5-10) bpm is acceptable, whereas a higher RR measurement accuracy is required when detecting neonatal apnea. Pressure-sensitive-mats (PSM) are non-invasive, contactless, and unobtrusive sensors that are well-suited to respiratory monitoring both at home and in hospitals. The use of PSM for respiratory monitoring in the adult patient population is gaining ground [2], [9]–[21]. There is, however, a lack of research on similar applications of the PSM technology in the infant and neonatal population. There are many potential advantages to adding an unobtrusive monitoring device such as the PSM to a neonate's hospital bed. It does not interrupt the routine care provided to the patient by the nurses and parents. It can act as a secondary estimator of RR during activities that require lead removal of standard monitoring devices. Lastly, the PSM is contactless, therefore, no adhesives are applied to the fragile and sensitive skin of a newborn. This paper explores the application of PSM for respiratory monitoring in the neonatal intensive care unit (NICU) of a hospital. To the best of our knowledge, this application is novel to the NICU. This paper extends our pilot study in [22], where we had compared time and frequency domain techniques to estimate a range of simulated neonatal RR by analyzing PSM data.

Considering the medical instrumentation and uncertainty measurement categories introduced by Parvis and Vallan [23], the PSM sensor falls in the first category, where the instrument is used to directly measure a patient's contact pressure data. Our previous research shows that drift and creep are factors that contribute to the uncertainty in the PSM measurements [24]. Drift is a random component of error that reflects a tendency of the system output to float higher or lower over continuous measurements caused by the fluctuation of the zero scale, whereas creep refers to the systematic trend that is observed in pressure values over time once the initial transient response has passed [25].

The Guide to the Expression of Uncertainty in Measurement (GUM) identifies that such random and systematic effects in the instrumentation give rise to measurement uncertainty that must be quantified [26]. Referring to section 3.2.2 in the GUM, random error is presumed to arise from stochastic temporal and spatial variations that influence the quantity that one is trying to measure [26]. Given enough observations, the expected value of these variations approaches zero. This is in alignment with our adopted definition of drift, where drift is the stochastic or random effect that causes the observed pressure values to float around the mean pressure value, such that the average drift value approaches zero.

S. Nizami, A. Bekele, M. Hozayen, and J.R. Green are with the Department of Systems and Computer Engineering, Carleton University, 1125 Colonel By Drive, Ottawa, ON K1S 5B6, Canada. (e-mail: {shermeen, jrgreen}@sce.carleton.ca).
K. Greenwood is with Clinical Engineering and J. Harrold is with Neonatology, Children's Hospital of Eastern Ontario, 401 Smyth Rd, Ottawa, ON K1H 8L1, Canada. (email: {kgreenwood, jharrold}@cheo.on.ca)

GUM section 3.2.2 further states that the experimental standard deviation of the average of a series of observations is "a measure of the uncertainty of the average due to random effects". We calculate the experimental standard deviation of the observed average pressure using Eq. (1) to quantify the uncertainty of the average pressure due the random effect of drift; thereby, denoting it as Drift (%).

Referring to section 3.2.3 in the GUM, systematic error arises from a recognized effect of an influence quantity on a measurement result. Given sufficient data, the systematic effect can be quantified, and unlike random effects, a correction factor can be applied to compensate for the systematic effect. In our recent research, we have quantified creep over long term experiments lasting over 14 hours and shown that it is a systemic effect [24].

We here: (a) quantify the standard uncertainty of the PSM data by calculating its standard deviation as percentage drift; (b) report the standard deviation of drift by using the moving block bootstrap method; and (c) compute the one-minute percent creep values. The present assessment differs from our previous work in that two clinical-grade patient simulators of differing size and mass are used with two different mattresses used in clinical practice.

The second category of medical instrumentation noted in Parvis and Vallan consists of systems and algorithms that convert patient data into derived features of clinical interest [23]. In this research, the development and evaluation of a modified frequency domain algorithm that estimates the RR from PSM data falls into this category. We conduct 28 experiments that examine five mixed effects to emulate the multiple or complex sources of variability present during real patient monitoring in an NICU. The mixed effects comprise one random effect, of the simulated RR value, and four fixed effects, which include internally and externally induced patient motion, two mattress types, grunting versus normal breathing patterns, supine and prone laying positions. We apply a previously developed frequency domain RR estimation algorithm [22] to these PSM data, as well as a modified RR estimation algorithm developed in this paper. The goal is to identify which of these effects significantly impact the accuracy of the two RR estimation algorithms. In accordance with the GUM [26], we determine the RR 95% confidence intervals to quantify the uncertainty in these estimates by performing a limits of agreement (LoA) analysis similar to Parker *et al.* [27]. Our research also complies with the Standards for Reporting Diagnostic Accuracy (STARD), where 95% confidence intervals are essential to determine the accuracy of patient-related diagnostic tests and devices [28].

The normative RR of an individual patient is dependent on many factors including, but not limited to, age, states of sleep or wakefulness, body mass index, and pathophysiology [29]. For a given patient, the clinician usually infers an expected RR value based on their expert knowledge and past experience. There is emerging research in observing the RR bounds of typically hospitalized patients, however, such research is patient-centric and not generalizable as shown by the comparison of three different RR scales in [30]. In addition, there is no mention of the required accuracy of these scales. Thus, we have chosen to adopt *a priori* the clinical rule that an RR accuracy within 5-10 bpm would be an acceptable limit of agreement, similar to [27].

In these experiments, patient motion is designated as the first mixed effect and includes both internally induced motion such as limb movement or seizures, and externally induced motion such as diaper changes, nasogastric tube insertion, or other routine or clinical interventions. Several groups have examined methods to detect and remediate motion artifact that is expected to negatively impact RR estimation. For example, in [31] the authors identify period of motion and censor recorded data during these periods when estimating RR over a longer period. Research in [32] develops methods to select the sensor with the best respiration signal in the presence of motion. Others have deployed an array of pressure sensors to detect the change in respiratory signal with respect to patient movement [13], [14], [33]. The second mixed effect is the type of mattress used under the patient. The effects of different mattress types for pressure distribution and patient support are well known [34], [35]. Estimates of RR are compared by placing the PSM on an overhead warmer bed mattress and a firmer crib mattress that are both routinely used in the NICU. As a third mixed effect, this research aims to distinguish between grunting and normal breathing patterns. Grunting is an important clinical symptom in the diagnosis of neonatal respiratory distress caused by lung abnormality [36], [37] or immaturity. Newborns presenting with neonatal respiratory distress are admitted to the NICU and treated with oxygen [38], [39] or positive pressure as needed. The fourth mixed effect is the laying position of the simulator, either supine or prone. Although it is now widely advised to maintain all neonates in the supine position, particularly at home for the prevention of Sudden Infant Death Syndrome, the prone position is occasionally used in the NICU while infants are closely monitored, as it can help with reflux and positioning of premature infants. Lastly, the fifth mixed effect is a set of breathing rates of 45, 60 and 75 breaths per minute (bpm). These RR values fall within the ranges observed in neonates, whether preterm or term born, as specified in [30], [40], [41]. This effect is random since the actual RR is typically unknown during actual patient monitoring and varies stochastically.

We analyze contact pressure data in the frequency domain to estimate the different RR by fast Fourier transformation (FFT), with subsequent identification of the frequency component that is contributing the largest signal power. Our previous research shows that frequency domain analysis of PSM data produces RR estimation results that are far superior to those obtained through time domain analysis [22].

The paper is organized as follows. The methods section describes: A. Research equipment; B. Measurements of the PSM; C. Mixed Effects Data Acquisition; D. Signal Pre-processing; E. Frequency domain analysis; F. Modified RR estimator; and G. Uncertainty Measurements in Respiratory Rate Estimation. This is followed by the results, discussion and conclusion sections.

## II. METHODS

### A. Research Equipment

The bench testing was conducted at the Children's Hospital of Eastern Ontario, Ottawa, Canada and Carleton University, Ottawa, Canada. The equipment included two neonatal patient simulators "SimNewB" and "Premature Anne" from Laerdal Medical Canada, Ltd., Toronto, Canada. "SimNewB" weighs 2790g (6.2 lb.) and is 51 cm (21 in.) long, these proportions are representative of a newborn baby [40][42]. Premature Anne is a realistically proportioned 25-week gestation, 900g (1.98 lb.) preterm mannequin developed in collaboration with the American Academy of Pediatrics (AAP). The two mattress types were a Giraffe overhead warmer neonatal bed mattress (GE Healthcare, USA) with a size of 65x48x4 cm (25.5L x 19W x 1.5D in.), and an open crib mattress that is significantly firmer and approximately double the size and depth of the former. The capacitive PSM used in the study is LX100:36.36.02 (XSensor Technology Corp. Calgary, Canada, XSensor.com). The experiments were conducted by placing the neonatal simulators on each of the two different mattresses.

The PSM sensor was placed on top of the mattress and covered with a sheet that is normally used in the NICU. The PSM sensor has a density of 1 sensel/in$^2$ with an overall sensing area of 18 x 18 in$^2$. The PSM connects to an X3 Pro Sensor Pack that feeds into an X3 Pro Electronic Platform that is connected via USB to a laptop running the X3 Pro software. The X3 Pro software was used to record PSM data and video simultaneously. Fig. 1 also shows the contact pressure image produced by the X3 Pro software in one frame during the acquisition of a supine dataset in the crib. The labels indicate the body parts of the simulator on the PSM. The shaded thorax area marks the region of interest for which the pressure data are analyzed for estimating the RR.

Acquired pressure values are calibrated by the X3 Pro software. A noise floor of 0.06 psi was used for the drift and creep calculations, such that sensel pressure values below this minimum value are excluded from the average pressure calculations.

*B. Measurements of the PSM*

To characterize the PSM measurements and the uncertainty in the measurements, we study the underlying random and systematic effects. The standard uncertainty of the PSM data is determined by calculating its standard deviation, which is reported as percentage drift using (1). Furthermore, we estimate the standard deviation of drift using moving block bootstrap method. For the moving block bootstrap method, a block size of 100 samples was applied where blocks are permitted to overlap. Bootstrap samples were assembled by drawing n/100 blocks with replacement, where the original record length was n samples. The standard deviation was calculated from 2000 bootstrap samples.

The one-minute percent creep represents a systemic effect on the PSM measurement; this is computed using (2). $P_n$ is the spatially averaged pressure across the entire mat in the *nth* time frame. $P_{avg}$ is the average of all $P_n$ values across *N* time frames. Drift is then simply the standard deviation of $P_n$ reported as a percentage of $P_{avg}$. Creep is the difference between the *Nth* pressure value $P_N$ and the first pressure value $P_1$, reported as a percentage of $P_{avg}$. Here $P_1$ and $P_N$ are also temporally averaged over five second windows to mitigate drift while estimating creep. To accommodate for different lengths of data, the creep is assumed to be linear over the first minute and is extrapolated over the first minute as in (2), where the length of the data segment in seconds equals *N* divided by the sampling rate, $f_s$.

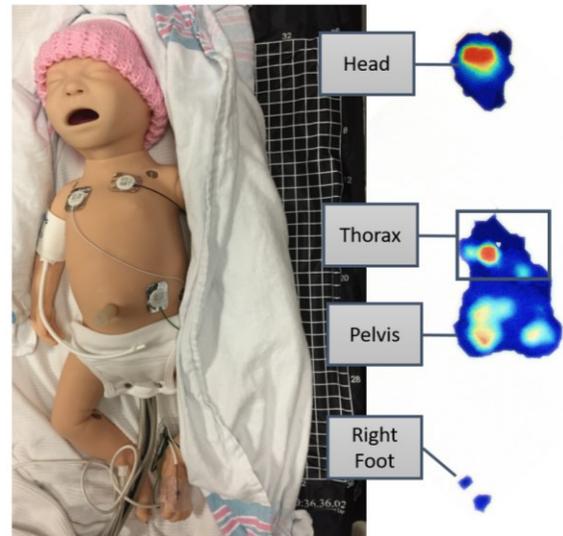

Fig. 1: SimNewB lying supine on XSensor PSM on the crib mattress, with its corresponding pressure image on the right

$$Drift\ (\%) = \left(\frac{\sqrt{\frac{\sum_{n=1}^{N}(P_n - P_{avg})^2}{N}}}{P_{avg}}\right) \times 100 \quad (1)$$

$$Creep\ (\%) = \frac{P_N - P_1}{P_{avg}} \times \frac{60 \times f_s}{N} \times 100 \quad (2)$$

In these experiments, the two neonatal models (SimNewB and Premature Anne) were placed on the PSM once on the overhead warmer mattress and once on the crib mattress, under the baseline conditions of no respiration and no motion. The first and last two seconds of data are discarded to discount any transient artifacts appearing in the pressure signal due to loading and off-loading. The values obtained by applying (1) and (2) are reported in Table I.

*C. Mixed Effects Model for RR Estimation*

A total of 28 trials were conducted representing various combinations of the five mixed effects. Of these experiments, 18 were on an overhead warmer while 10 were on the crib mattress; 19 represented normal breathing patterns while 9 represented breathing with grunting; in 16 experiments the simulator was lying in the supine position with the remaining 12 in the prone position; and 6 trials were simulated at an RR of 45 bpm, 15 at 60 bpm, and 7 at 75 bpm. Breathing is a mechanical function of the simulator, where air from an external compressor is used to cyclically inflate an air sac simulating both lungs. Average contact pressure data acquired from the thorax region were analyzed to extract breathing patterns. The trials were 30-80 seconds long; contact pressure data were acquired at a sampling rate of 20 frames/sec at a noise floor of 0.097 psi.

*D. Signal Pre-processing*

The PSM data are normalized to remove DC bias similar to the method in [43]. The DC bias is caused by static forces from the load placed on the PSM; static loading is irrelevant when estimating RR. To normalize, the average of all data points in the analysis window is calculated and subtracted from each data point in the window. The DC signal causes a very large peak at zero Hz in the periodogram, thus overshadowing the power of the fundamental frequency of the respiratory cycle. Therefore, it is necessary to normalize the average contact pressure data.

*E. Frequency domain analysis*

For infants with a corrected age in the range of 1 to 79 weeks, the respiratory signal lies in the low-frequency band [41]. In this research, the random effect was provided by setting the neonatal simulator's RR at three different rates of 45, 60 and 75 bpm, corresponding to frequencies of 0.75-1.25 Hz. The RR estimation method is implemented as follows: first, MATLAB's *fft* function is applied to the signal over windows of 20s with a 50% overlap. Then, for each window, the frequency peak with the highest power contribution is identified. Finally, the mean of the peak frequencies is multiplied by 60 to estimate the RR in bpm. The window size of 20s and 50% stride were selected arbitrarily and were not optimized. This procedure was applied to each dataset in their entirety. However, this procedure can also be implemented using a sliding window approach for providing real-time RR estimates.

*F. Modified RR estimator*

Using the method described above leads to the inference that motion poses the greatest challenge in the estimation of RR, as discussed in the Results and Discussion sections. To overcome this limitation, we present here a modified RR estimator that suppresses motion artifacts and isolates the breathing signal prior to applying the frequency-domain RR estimation. Movement in these experiments results from either patient-induced (mechanical) movement or externally-induced (simulated interventions). A frequency-domain analysis of the resulting motion artifacts indicated that these artifacts were primarily low frequency, partially overlapping with the breathing signal. It is known that spectral peak search is vulnerable to such additive noise. To quantify this limitation, we know that (a) the respiratory signal is band-limited, and (b) we know the respiration signal frequency (and its harmonics). Thus, we compute the power within the signal band versus all other frequencies to obtain the signal-to-noise ratio (SNR). The average SNR was -2.87 dB across all the trials with motion, and it was 3.71 dB for all the trials without motion. Due to the limitation that noise may overlap in frequency with the signal, this method may slightly overestimate the SNR.

As a first step, a moving average filter is applied to the raw data to extract an estimate of the low frequency motion artifact. The window width of the filter is heuristically identified as 1.5 sec or 30 frames in this case. If a smaller window is used, the RR signal will be compromised. If the window is too large, then the motion signal will not be isolated for subsequent subtraction. This smoothing process is depicted in Fig. 2, where the top graph shows the raw data signal of average pressure (psi) obtained from the PSM. The middle graph shows the result of the moving average filter, representing the motion artifact. The graph at the bottom shows the signal obtained by subtracting the smoothed signal from the raw signal. The low frequency signal noise due to motion is removed and the higher frequency breathing signal is apparent. This is followed by normalization as described in II.D and the *fft* analysis in II.E.

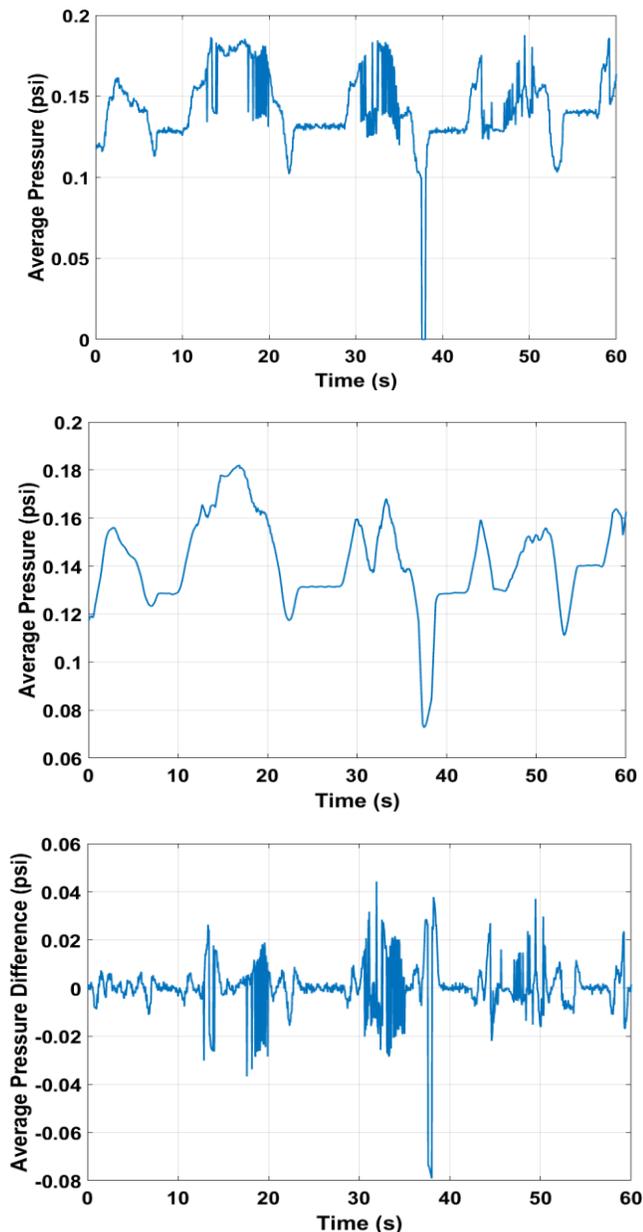

Fig. 2: Modified RR estimator method. Top: Raw pressure signal. Middle: Smoothed signal after application of the moving average filter, representing isolated motion artifact. Bottom: Resulting signal ready for RR estimation after subtracting the smoothed signal from the raw signal.

*G. Uncertainty Measurement in Respiratory Rate Estimation*

In terms of quantifying the uncertainty of both RR estimation methods, we follow the GUM to provide the 95% confidence intervals on the error in estimated RR, through a limits of agreement analysis. Following the approach from [27], we applied a methodology of mixed effects LoA to obtain a 95% confidence interval for RR estimation error. To calculate the limits of agreement for each RR estimator, we analyze the paired difference between the RR estimate and the gold standard RR using a mixed effects regression model. Here, the simulated value of 45, 60 or 75 bpm in any given experiment is used to identify a patient and is treated as a random effect.

Patient movement, mattress type, grunting, and position are all expected to have a systematic and predictable influence on the RR estimation, thus they are included as a fixed effect. The mixed effects model is defined as follows by (3):

$$Y_m = X\beta + Zb + \varepsilon \qquad (3)$$

For a total of m = 2 methods, N = 28 experiments, p = 4 fixed effects and q = 1 random effects: $Y_m$ is an N x 1 vector of paired differences for method m, $X$ is an N x p matrix of fixed effect values, $\beta$ is a p x 1 a vector of fixed effects, $Z$ is an N x q matrix of random effect values, b is a q x 1 vector of random effects, and $\varepsilon$ is an N x 1 error vector.

The assumptions for the linear mixed-effects model is that random effects vector, b, and the error vector, ε, are normally distributed and independent from each other. The 95% LoA is calculated as follows:
1. Fit the mixed effects model and extract the between-patient variance (V1) and within-patient variance (V2); V2 accounts for all fixed effects as combined residuals.
2. To estimate the standard deviation (SD), combine the variance with a simple sum formula (SD = $\sqrt{V_1 + V_2}$)
3. Fit a second random effects model that excludes the fixed effects to estimate the mean bias (m)
4. Calculate the 95% LoA as $[m - 1.96\ x\ SD, m + 1.96\ x\ SD]$

Next, further analysis is done to quantify the significance of each of the four fixed effects: motion, mattress type, grunting and position. For each fixed effect, a new model is constructed by excluding that effect from the model to obtain a new estimate for the mean bias and 95% LoA. For 95% confidence interval analysis, we assume that the paired differences will be normally distributed over the selected range of simulated gold standard RR values. Then, the new model is compared with the model that contains all the effects, through a likelihood ratio test, as described in [44]. In addition, we compute the Pearson correlation coefficient between the RR estimates and the gold standard RR values for all 28 experiments for both RR estimation methods.

### III. RESULTS

Table I shows the average contact pressure measurement (psi), average percent contact area, percent drift, one-minute percent creep, and standard deviation of drift values for the two neonatal models and mattress types. For example, the

TABLE I
CHARACTERISTICS OF THE XSENSOR PSM

|  | Premature Anne | | SimNewB | |
| --- | --- | --- | --- | --- |
|  | Crib | Overhead Warmer | Crib | Overhead Warmer |
| $P_{avg}$ (psi) | 0.136 | 0.099 | 0.139 | 0.101 |
| Avg contact area (%) | 3.086 | 5.093 | 8.873 | 15.201 |
| Creep (%) | -0.097 | 0.091 | 0.341 | 0.228 |
| Drift (%) | 0.586 | 0.213 | 0.163 | 0.093 |
| Std of Drift (%) | 0.008 | 0.009 | 0.018 | 0.009 |

TABLE II
95% LoA RESULTS FOR ORIGINAL RR ESTIMATION AND MODIFIED METHOD

| Method | Mean bias (Fixed effect 95%LoA) |
| --- | --- |
| *RR* | 5.38 (3.64 to 7.12) |
| *Modified RR* | 0.56 (-2.26 to 3.37) |

TABLE III
LIKELIHOOD RATIO TEST RESULTS FOR RR METHOD

| Excluded effect | Mean bias (Fixed effect 95% LoA) | Likelihood ratio test results |
| --- | --- | --- |
| *Motion* | 5.38 (-24.00 to 34.76) | ($\chi^2$ (2) = 159.26, p = 0) |
| *Mattress Type* | 7.49 (5.96 to 9.04) | ($\chi^2$ (1) = 0.4693, p = 0.49) |
| *Grunting* | 7.50 (5.97 to 9.03) | ($\chi^2$ (1) = 0.0237, p = 0.88) |
| *Position* | 7.50 (5.97 to 9.03) | ($\chi^2$ (1) = 0, p = 1.00) |

TABLE IV
95% LoA RESULT WHEN MOTION IS EXCLUDED FROM FIXED EFFECTS

| Method | Mean bias (Fixed effect 95% LoA) | Likelihood test results |
| --- | --- | --- |
| *RR* | 5.38 (-24.00 to 34.76) | ($\chi^2$ (2) = 159.26, p = 0) |
| *Modified RR* | 0.56 (-2.52 to 3.64) | ($\chi^2$ (2) = 5.18, p = 0.075) |

Premature Anne model exerts an average pressure of 0.136 psi over 3.086% of the total PSM area on the crib mattress, and an average pressure of only 0.099 psi over 5.093% of total PSM area when using the overhead warmer mattress. Standard uncertainty in the average contact pressure measurement is reported as percent drift in Table I.

Table II shows the mean bias and 95% LoA for the two RR estimation methods. In Tables III-IV, the reported results are chi-square statistic at the specified degrees of freedom $\chi^2$ *(DF)*, *p* values for test of significance, and the *Lower* and *Upper* 95% limits of agreement. In Table III, results from 95% LoA and likelihood ratio test for the model constructed by excluding each one of the fixed effects is presented for the original RR estimation method. As can be seen, only the fixed effect of motion has a significant impact on RR estimation error, with a p-value of 0. This motivated the development of a modified RR estimation method that specifically mitigates motion artifact prior to estimating RR from the PSM data. Table IV shows the likelihood ratio test results for the model excluding motion effect specifically for both RR methods. The mean bias due to motion has been reduced from 5.38 to 0.56 bpm and the effect of motion no longer induces a statistically significant effect on RR estimation error (p-value increases from 0 to 0.075) [44]. The Pearson correlation coefficient is 0.53 for the first RR estimation method and 0.99 for the second modified RR estimation method.

### IV. DISCUSSION

As observed from Table I, each neonatal model exerts different pressures on different surfaces. The crib mattress is firmer than the overhead warmer mattress so the recorded contact pressures for both models are greater on the crib mattress. As expected, the softness and greater flexibility of the overhead warmer mattress causes the applied load to be distributed over a larger area, hence resulting in a slightly reduced average pressure. Also, the larger mass and size of SimNewB covers a larger area and exerts greater average pressure on both mattress types. We see an increase in drift when using this Premature Anne model when compared with the larger SimNewB doll. This result is not unexpected due to the smaller contact area, now reported in Table I, which results in fewer active sensels in the PSM. Since the contact pressure is measured as the spatial average across all active sensels, then the Premature Anne generates a noisier estimate of contact pressure due to the smaller number of active sensels.

When comparing mattress types, the drift values show a consistent decrease from the crib to the overhead warmer mattress. This is consistent with the drop in average pressure as described above, since the drift is measured as a percentage of average applied pressure. For the smaller Premature Anne model, the rate of creep is essentially negligible, and the measured contact pressure is dominated by drift. This results in an apparently negative creep on the crib mattress. For short-term experiments such as these, the uncertainty is dominated by drift, whereas for longer term experiments, one would expect creep to be the dominant effect. Long-term experiments should be conducted for evaluating drift and creep when the PSM is being considered for continuous patient monitoring, as done in [24].

It is observed from the results in Table II that, for the original RR method, the mean bias was found to be 5.38 bpm with 95% LoA of [3.64, 7.12] bpm. This result is acceptable for current clinical practice, but it doesn't fall within ± 5 bpm which is required for use in monitoring apnea. While inspecting the contribution of each of the fixed effects, as per the results in Table III, it can be observed that motion significantly affected the RR estimate ($\chi^2$ (2) = 159.26, p = 0), widening the 95% LoA from [3.64, 7.12] bpm to [-24.00, 34.76] bpm. The negative lower limit indicates underestimation of the RR. It is also important to note that this new LoA doesn't fall within the +/- 10 bpm requirement. The sources of motion in this study were highly heterogeneous, including internally induced motions such as limb movement or seizures, and externally induced motions such as diaper changes, nasal gavage tube insertion, and other routine and clinical interventions. Some of these types of motion more severely affect our ability to recover RR than others. This results in the wide confidence intervals observed. Given more data, it would be possible to analyze each motion type separately, likely resulting in tighter confidence intervals for each type of motion.

The modified RR estimation method, in comparison with the gold standard RR values, has a mean bias of 0.56 bpm and 95% LoA [-2.26, 3.37] bpm. This meets the more favorable ±5 bpm requirement. Moreover, from Table IV, it can be seen that the modified RR estimation method is not significantly sensitive to motion when compared to the original RR method.

The *p* values for the other three fixed effects of grunting (*p=0.88*), mattress type (*p=0.49*), and position (*p=1.0*) are greater than 0.05. This is in agreement with our previous work in [22] which showed that the original RR method performs accurately regardless of grunting, mattress type, and position when estimating RR in the absence of motion. This indicates that these factors do not seem to have a significant impact on our ability to estimate RR using either method.

## V. Conclusion

PSM use is advantageous in the population of critically ill babies as the mats are non-invasive, non-contact, and unobtrusive. This research characterized the XSensor LX100:36.36.02 PSM in terms of contact pressure measurements and the associated measurement uncertainty, which are essentially dependent on the applied load and contact surface. The experiments were conducted using clinically relevant neonatal simulator models and mattresses to emulate the PSM characteristics within a real NICU setting. The two models were selected to represent the expected patient sizes and masses within the NICU population.

A mixed effects analysis was conducted to examine the impact of five experimental variables typical of clinical environments: mattress type, patient movement, actual RR, patient position, and grunting during breathing. Of these five, only the patient movement was found to significantly impact our ability to estimate RR. Therefore, a modified RR estimator is proposed, where movement artifact is first identified and removed in the time domain, prior to frequency-domain estimation of the dominant frequency corresponding to the RR. Evaluation of this modified estimator indicates that it is superior in the presence of movement. Future work will examine incorporating movement detection algorithms such that remediation is only applied during periods of actual patient motion to preserve accuracy during periods of no motion.

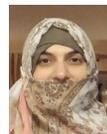

**Shermeen Nizami** (S'10–M'17) received the B.A.Sc. degree in Electrical Engineering from the University of Engineering and Technology, Lahore, Pakistan and the M.A.Sc. and Ph.D. degrees in Electrical and Computer Engineering from Carleton University, Ottawa, ON, Canada, in 1996, 2004, and 2016, respectively.

She is currently a Postdoctoral Research Fellow cross-appointed at the IBM Centre for Advanced Studies, Ottawa, ON and the Department of Systems and Computer Engineering, Carleton University, Ottawa, ON. Her current research interests include neonatal patient monitoring, artifact detection and biomedical instrumentation.

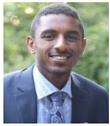
**Amente Bekele** received the B.Eng. degree in Computer Systems Engineering from Carleton University, Ottawa, ON, Canada in 2016.

He is currently pursuing an M.A.Sc. degree in Electrical and Computer Engineering with specialization in Data Science at Carleton University Ottawa, ON, Canada.

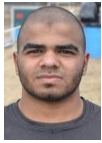
**Mohamed Hozayen** is currently pursuing a B.Eng. in Computer Systems Engineering at Carleton University, Ottawa, ON.

His research includes ECG biometric identification and developing software for exporting and parsing patient monitor data.

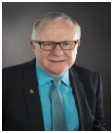
**Kimberley J. Greenwood** received his diploma in Electrical Engineering Technology (1984) from Ryerson Polytechnical Institute, Toronto, ON, Canada. He received a B.A.Sc. (2006) in Technology Management from Bemidji State University, Bemidji, Minnesota, USA and an M.A.Sc. in Biomedical Engineering from Carleton University (2010), Ottawa, Ontario, Canada. He is a licensed professional engineer in the Province of Ontario and is a certified clinical engineer.

He is currently the Director of Clinical Engineering, Facilities Management and Planning at the Children's Hospital of Eastern Ontario of Ottawa, ON, Canada.

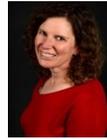
**JoAnn Harrold** received the BSc (1994) and MD (1997) degrees from McMaster University, Hamilton, ON, Canada. She holds Royal College certification in Pediatrics (training at University of Toronto, Toronto, ON, Canada) and in Neonatal-Perinatal Medicine (training at McMaster University, Hamilton, ON, Canada.

She is currently an Associate Professor in the Faculty of Medicine at University of Ottawa, Division Chief of Neonatology at Children's Hospital of Eastern Ontario and Division Head of Newborn Care at The Ottawa Hospital, Ottawa, ON, Canada.

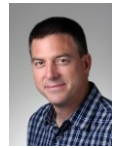
**James R. Green** (S'02–M'05–SM'10) received the B.A.Sc. degree in Systems Design Engineering (1998) from the University of Waterloo, Waterloo, ON, Canada, and M.A.Sc. (2000) and Ph.D. (2005) degrees in Electrical and Computer Engineering from Queen's University, Kingston, ON, Canada.

He is currently an Associate Professor with the Department of Systems and Computer Engineering, Carleton University, Ottawa, ON. His research interests include pattern classification challenges within biomedical informatics, patient monitoring, computational acceleration of scientific computing, and the design of novel assistive devices.